\documentclass[sigconf,screen]{acmart}

\usepackage{multirow}
\usepackage{bm}
\usepackage[justification=centering]{caption}
\usepackage{balance}
\usepackage{float}
\usepackage{tabularx}
\usepackage{longtable}
\usepackage{hyperref}
\usepackage{graphicx} 
\usepackage{subfigure} 
\usepackage{indentfirst} 
\usepackage{listings}
\usepackage{afterpage}
\usepackage{verbatim}
\usepackage{makecell}
\usepackage{colortbl}

\lstdefinelanguage{Kotlin}{
  morekeywords={val, var, fun, if, else, while, for, in, return, open, class},
  sensitive=true,
  morecomment=[l]{//},
  morecomment=[s]{/*}{*/},
  morestring=[b]",
  morestring=[s]{"""*}{*"""},
}

\definecolor{codegreen}{rgb}{0,0.6,0}
\definecolor{codegray}{rgb}{0.5,0.5,0.5}
\definecolor{codepurple}{rgb}{0.58,0,0.82}
\definecolor{backcolour}{rgb}{0.95,0.95,0.92}

\setcopyright{acmcopyright}
\copyrightyear{2024}
\acmYear{2024}
\acmDOI{10.1145/xxxxxxx.xxxxxxx}
\acmConference[ASE '24]{ASE'24: the 39th IEEE/ACM International Conference on Automated Software Engineering (ASE '24)}{October 27 -- November 01, 2024}{Sacramento, California, USA}
\acmBooktitle{Proceedings of the 39th IEEE/ACM International Conference on Automated Software Engineering (ASE '24), October 27 -- November 01, 2024, Sacramento, California, USA}
\acmISBN{978-1-4503-XXXX-X/18/06}

\begin{document}

\title{Depends-Kotlin: A Cross-Language Kotlin Dependency Extractor}

\author{Qiong Feng, Xiaotian Ma, Huan Ji, Wei Song}
\affiliation{%
  \institution{Nanjing University of Science and Technology}
  \city{Nanjing}
  \country{China}
}
\email{{qiongfeng, xyzboom, alex, wsong}@njust.edu.cn}

\author{Peng Liang}
\authornote{This work is partially funded by NSFC with No. 62172311.}
\affiliation{%
  \institution{School of Computer Science, Wuhan University}
  \city{Wuhan}
  \country{China}
}
\email{liangp@whu.edu.cn}

\renewcommand{\shortauthors}{Feng et al.}

\begin{abstract}
\noindent Since Google introduced Kotlin as an official programming language for developing Android apps in 2017, Kotlin has gained widespread adoption in Android development. However, compared to Java, there is limited support for Kotlin code dependency analysis, which is the foundation to software analysis. To bridge this gap, we develop \textit{Depends-Kotlin} to extract entities and their dependencies in Kotlin source code. Not only does \textit{Depends-Kotlin} support extracting entities' dependencies in Kotlin code, but it can also extract dependency relations between Kotlin and Java. Using three open-source Kotlin-Java mixing projects as our subjects, \textit{Depends-Kotlin} demonstrates high accuracy and performance in resolving Kotlin-Kotlin and Kotlin-Java dependencies relations. The source code of \textit{Depends-Kotlin} and the dataset used have been made available at \url{https://github.com/XYZboom/depends-kotlin}. We also provide a screencast presenting \textit{Depends-Kotlin} at \url{https://youtu.be/ZPq8SRhgXzM}.
\end{abstract}

%to  a Java project with confirmed dependencies as a benchmark and converted this project to two projects: Kotlin-only and a combination of Kotlin and Java. The dependencies in these two projects were then extracted using our tool. The consistency observed among dependency relations in all three projects confirms the accuracy of \textit{Depends-Kotlin}. Furthermore, the performance of \textit{Depends-Kotlin} was assessed using another three projects of varying sizes.

\maketitle

\section{Introduction}
\label{intro}

The dependency relations among entities in the source code form the foundations of software architecture analysis~\cite{mo:2019tse,cui:2019icse,xiao:2021tse}. Since Google introduced Kotlin as an official programming language for developing Android apps in 2017, Kotlin has gained widespread adoption in Android development. According to recent empirical studies, a large number of Android apps have been continuously migrated from Java to Kotlin~\cite{ardito2020effectiveness, mateus2020adoption}. However, compared to dependency analysis tools that support Java, such tool support for Kotlin is limited. To our knowledge, many well-known dependency analysis tools that support the Java language, such as Understand~\cite{understand}, and DV8~\cite{dv8}, currently do not offer support for the Kotlin language.

Kotlin dependency resolution faces two \textbf{challenges}. First, Kotlin is known as ``\textit{concise, expressive, and designed to be type and null-safe}''. It contains a lot of syntactic sugar to ensure these features, thereby increasing the difficulty of resolving entity types and dependencies. Consider the example in Listing 1: Line 1 declares a class named \texttt{Bar} with a property \texttt{x} of type \texttt{Int} (property in Kotlin is similar to field in Java). Line 3 declares a high-level function named \texttt{calculate} that takes a lambda expression as its parameter. As shown in this line, the lambda takes in a \texttt{Bar} type (called receiver type in Kotlin), and returns an \texttt{Int}. Line 5 declares another class named \texttt{Foo}, also with a property \texttt{x} of type \texttt{Int}. Line 6 declares \texttt{calculateInFoo}, which is a member function of the \texttt{Foo} class. It invokes the \texttt{calculate} function in Line 7. Since the lambda parameter takes type \texttt{Bar}, the \texttt{add} function invoked in the lambda accesses the \texttt{x} property of the current \texttt{Bar} instance, not the \texttt{Foo} instance. The use of lambdas with receiver types as input allows concise syntax but it increases the difficulty of dependency resolution. In order to resolve this dependency (\texttt{calculateInFoo} \textbf{use} \texttt{Bar.x}), we need to locate \texttt{calculate}'s parameters and further trace them down to the correct type and properties. Furthermore, Kotlin has its own dependency types, such as \textbf{delegate} and \textbf{extension}, which lacks support from existing tools..

\begin{minipage}[c]{0.48\textwidth}
\begin{lstlisting}[language=Kotlin, caption=Kotlin syntax sugar example]
class Bar(val x: Int)
// Declaration of class Bar with the property x
fun calculate(param: Bar.() -> Int) {}
/* Function 'calculate' takes a lambda with a receiver type Bar as its parameter.*/
class Foo(val x: Int) {
    fun calculateInFoo() {
        calculate { add(x) } 
        // x in add(x) here is actually Bar.x
    }
}
\end{lstlisting}
\end{minipage}

%// Declaration of class Bar with the property x
%/* Function 'calculate' takes a lambda with a receiver type Bar as its parameter.*/

Second, Kotlin is designed to be fully interoperable with Java and can run on the JVM, which makes it easy to continuously migrate Java code to Kotlin. Listing 2 includes a Kotlin class \texttt{BarKotlin} and a Java class \texttt{FooJava}.  Lines 3-5 demonstrate that the Java class \texttt{FooJava} accepts and interacts with the Kotlin class \texttt{BarKotlin} by invoking a \texttt{getX()} method of the Kotlin class. However, in the Kotlin code the getter method is implicitly generated by JVM for the class property. If we analyze only static code, such dependencies cannot be recognized. Since a significant percentage of Android apps involve both Java and Kotlin code~\cite{ardito2020effectiveness, mateus2020adoption}, a tool for resolving Kotlin dependencies should not only address dependency relations in Kotlin but also handle dependencies between Java and Kotlin.

% from a Kotlin class to a Java method 
%, there is no explicit \texttt{getX()} method, as 

\begin{minipage}[c]{0.48\textwidth}
\begin{lstlisting}[language=Kotlin, caption=Java-Kotlin implicit invocation]
class BarKotlin(val x: Int)
public class FooJava {
    public static void func(BarKotlin bar) {
        System.out.println(bar.getX());
    }
}
\end{lstlisting}
\end{minipage}

To tackle these two challenges, we propose \textit{Depends-Kotlin}, a cross-language Kotlin dependency extractor. \textit{Depends-Kotlin} is based on the \textit{Depends} framework, an open-source project designed for code dependency analysis~\cite{jin2019enre}. We enhance \textit{Depends} to support the Kotlin language by performing multi-round inference for getting Kotlin's specific types and dependency relations. Additionally, we address cross-dependency relations between Kotlin and Java by refactoring the architecture of the original \textit{Depends} framework and filling in the implicit code of Kotlin properties, which is handled by JVM and not shown in the source code. \textit{Depends-Kotlin} is evaluated on three open-source projects containing both Kotlin and Java code and demonstrates its ability to accurately and efficiently extract Kotlin-Java and Kotlin-Kotlin dependencies.

\begin{figure}[!tp]
    \centering
    \includegraphics[width=0.61\columnwidth]{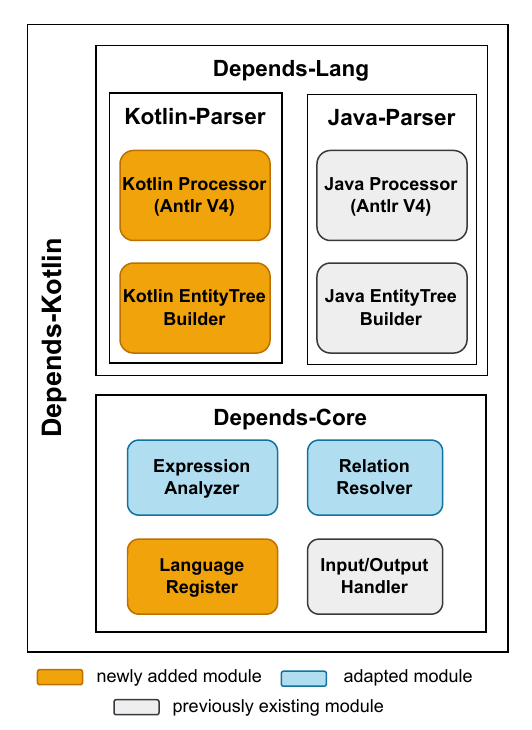}
    \vspace{-4mm}
    \caption{Plug-in architecture of \textit{Depends-Kotlin}}
    \label{fig:framework}
    \vspace{-6mm}
\end{figure}

\section{Methodology}
\label{sec2}

\subsection{Framework}

Figure~\ref{fig:framework} shows the architecture of \textit{Depends-Kotlin}. We adopted a plug-in architecture~\cite{wolfinger:2006component} and decomposed the original \textit{Depends} architecture into two parts: \textit{Depends-Core} (the main framework) and \textit{Depends-Lang} (plug-ins). \textit{Depends-Lang} manages the entity parsing for specific languages, while \textit{Depends-Core} handles the resolution of dependencies for the parsed entities. This architecture refactoring decision aims to enable the handling of cross-language dependencies, which is not supported by the original \textit{Depends} architecture. To this end, we added in \textit{Depends-Core} a \texttt{Language Register} module, which leverages Java Service Provider Interface (SPI) mechanism and allows registering different programming languages simultaneously. For a specific language, its language processor needs to extend the interface and has been registered in the SPI file. We also modified the existing \texttt{Relation Resolver} module in \textit{Depends-Core} to generate the implicit code, specifically the getter/setter code for Kotlin properties mentioned in Section ~\ref{intro}. This modification helps the detection of Java-Kotlin interactions.

In order to process Kotlin's unique syntax and resolve specific dependency relations, we modified the existing \texttt{Expression Analyser} and \texttt{Relation Resolver} modules in \textit{Depends-Core}. Specifically, We analyzed Kotlin's \textbf{extension} functions and properties in \texttt{Expression Analyser}. We also identified the scope of Kotlin's functions by adding related code as the context in \texttt{Expression Analyser}. Furthermore, we adapted \texttt{Relation Resolver} to handle dependencies with built-in types and new unique dependency relations in Kotlin. For example, as method members of built-in types in Java always return built-in types, the original \textit{Depends} chose to focus on the dependencies of analyzed source files and ignored built-in types. However, due to Kotlin extension functions, built-in types can return any types. This can greatly impact dependencies in the analyzed source code. Consequently we modified the \texttt{Relation Resolver} module to resolve such \textbf{extension} dependency relation.

For Kotlin's entity parsing, we adopted the same logic of Java-parser in the original \textit{Depends} framework. We used \textit{Antlr v4} to generate an Abstract Syntax Tree (AST) parser from a Kotlin grammar, which is provided on the official Kotlin website. This module is named as \texttt{Kotlin Processor} as shown in Figure~\ref{fig:framework}. Subsequently, the  \texttt{Kotlin EntityTree Builder} module store parsed AST nodes into concrete entities, such as files, packages, types, expressions, functions, and properties. In this step, some types in expressions can be resolved in advance, which can reduce subsequent workload. As previously mentioned, the parser module of a specific language can work as a plug-in to the \textit{Depends-Core} framework, enabling the analysis of other languages in the future.

\subsection{Entity and Dependency Relation Resolution}

\begin{table}[!tbp]
	\centering
        \small
        \caption{Dependency Relations supported by \textit{Depends-Kotlin}}
	\label{tbl:dp_types}
 \vspace{-4mm}
\begin{tabular}{c|l}
\toprule
\textbf{Relation} & \textbf{Description}                                    \\ \hline  
{\sffamily Import}   & a file imports another class, enum, static method    \\ \hline
{\sffamily Contain}   & a class holds another class's object as field
\\ \hline
{\sffamily Extend}   & a class extends a parent class        
\\ \hline
{\sffamily Implement}   & a class implements an interface     
\\ \hline
{\sffamily Call}     & an expression invokes another method       \\ \hline
{\sffamily Create}     & an expression in a method create an object
\\ \hline
{\sffamily Cast}     & an expression does a cast to a type
\\ \hline
{\sffamily Annotation}     & an entity uses an annotation
\\ \hline
{\sffamily Use}      & a method access a variable in its scope              \\ \hline
{\sffamily Parameter}& a method use another type as its parameter           \\ \hline
{\sffamily Return}   & a method returns another type                        \\ \hline
{\sffamily Delegate}   & a class delegates another class                    \\ \hline
{\sffamily Extension}   & a method is an extension of a class               \\        \bottomrule
\end{tabular}  
\\
{\raggedright Note: The first 11 dependency relations can be observed between Java and Kotlin entities. The last two are exclusive to Kotlin-Kotlin and Kotlin-Java.  \par}
\vspace{-6mm}
\end{table}

\begin{table*}
	\begin{minipage}{0.7\linewidth}
		\caption{Dependency Relations Extracted by \textit{Depends-Kotlin}}
		\label{tbl:extracted-deps}
		\centering
		\small
  \vspace{-4mm}
    \begin{tabular}{c|cccc|cccc|cccc||c}
    \hline
       \multirow{2}{*}{}  & \multicolumn{4}{|c|}{\textbf{DSU-Sideloader}}  & \multicolumn{4}{|c|}{\textbf{Flap}} & \multicolumn{4}{|c||}{\textbf{RootEncoder}} & \multirow{2}{*}{\textbf{Total}}   \\ \cline{2-13}
         & K-J & J-K & K-K & J-J & K-J & J-K & K-K & J-J & K-J & J-K & K-K & J-J  \\ \hline
        Import & 19 & 0 & 170 & 1 & 3 & 0 & 170 & 1 & 102 & 190 & 913 & 193 & \cellcolor{gray!30} 1762 \\ \hline
        Contain & 6 & 0 & 34 & 5 & 4 & 0 & 46 & 63 & 10 & 256 & 318 & 73  & 815 \\ \hline
        Extend & 1 & 0 & 4 & 0 & 0 & 0 & 20 & 2 & 6 & 1 & 101 & 75  & 210 \\ \hline
        Implement & 1 & 0 & 0 & 7 & 0 & 0 & 36 & 1 & 9 & 4 & 8 & 3 & 69 \\ \hline
        Call & 50 & 0 & 450 & 81 & 3 & 0 & 370 & 479 & 62 & 423 & 1278 & 1316 & \cellcolor{gray!30} 4512 \\ \hline
        Create & 1 & 0 & 31 & 2 & 3 & 0 & 94 & 25 & 59 & 71 & 489 & 57 & 832 \\ \hline
        Cast & 0 & 0 & 0 & 1 & 5 & 0 & 10 & 15 & 0 & 8 & 48 & 9 & 96 \\ \hline
        Annotation & 0 & 0 & 0 & 0 & 0 & 0 & 0 & 2 & 0 & 0 & 0 & 0 & 2 \\ \hline
        Use & 70 & 0 & 703 & 187 & 9 & 0 & 713 & 1093 & 141 & 203 & 3331 & 4317 & \cellcolor{gray!30} 10767  \\ \hline
        Parameter & 1 & 0 & 48 & 9 & 0 & 0 & 40 & 46 & 29 & 121 & 238 & 145  & 677 \\ \hline
        Return & 3 & 0 & 36 & 16 & 0 & 0 & 71 & 24 & 2 & 30 & 263 & 30 & 475 \\ \hline
        Delegate & 0 & - & 0 & - & 0 & - & 2 & - & 0 & - & 0 & - & \cellcolor{gray!30} 2 \\ \hline
        Extension & 0 & - & 0 & - & 0 & - & 6 & - & 0 & - & 3 & - & \cellcolor{gray!30} 9 \\ \hline \hline
        \textbf{Total} & \cellcolor{orange!50}152 & 0 & \cellcolor{gray!30} 1476 & \cellcolor{gray!30}  309 & \cellcolor{orange!50}27 & 0 & 1578 & 1751 & \cellcolor{orange!50} 420 & \cellcolor{green!40} 1307 & \cellcolor{gray!30} 6990 & \cellcolor{gray!30} 6218 \\ \hline
    \end{tabular}
    \vspace{-6mm}
	\end{minipage}\hfill
	\begin{minipage}{0.25\linewidth}
		\centering
		\includegraphics[width=0.5\linewidth,height=5cm]{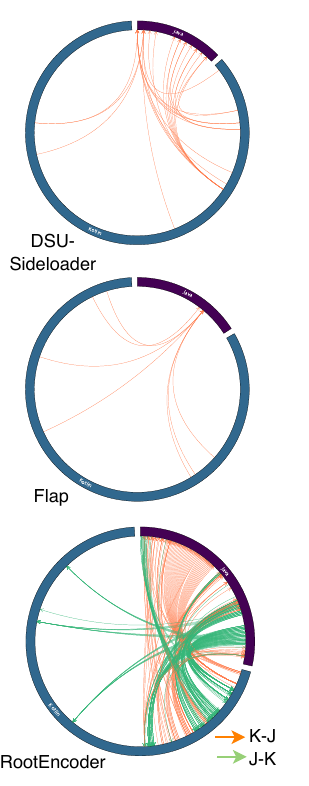}
		\captionof{figure}{J-K/K-J Dependency visualization(\textcolor{orange}{orange} denotes K depends on J; \textcolor{green}{green} denotes J depends on K)}
		\label{fig:visual}
\vspace{-6mm}
	\end{minipage}
\end{table*}

We followed the same logic of the original \textit{Depends} framework and performed multi-round inference for getting entities and dependency relations. An entity is represented by a tuple (\texttt{id}, \texttt{name}, \texttt{entityType}, \texttt{context}) and a dependency relation is represented by (\texttt{sourceId}, \texttt{targetId}, \texttt{dependencyType}, \texttt{weight}). For simple dependency relations, such as {\sffamily extend}, function {\sffamily parameter}, function {\sffamily return} and {\sffamily delegate}, such dependency relations between two entities can be collected directly when analyzing expressions. For dependency relations, such as member {\sffamily access} and member function {\sffamily call} on expressions of unknown types, we conducted type inference. The basic idea is to deduct types from what is known to what is unknown. Any operation on an expression with a known type, such as a member function call, will yield the next expression with a known type. This process is recursively repeated until the entire expression's type inference is completed. 

An additional search is required to resolve Kotlin {\sffamily extension} functions during this process, as the scope of the expression containing {\sffamily extension} functions needs to be identified. While traversing the syntax tree, if a function has a receiver type, it is marked as an extension function and the extended type is the receiver type. After the type system processing is complete, extension functions can be located based on their marks during the traversal of the entity tree. The extended types and the extension relationship of the function can then be recorded.

Our tool framework also automatically generates implicit code for Kotlin, and it distinguishes Java and Kotlin entities by assigning them different labels. When resolving expressions involving Kotlin-Java interactions, such cross-language relations can be easily detected. Table~\ref{tbl:dp_types} lists the dependency relations currently supported by \textit{Depends-Kotlin}. Take Listing 2 for example, a Java expression is calling a Kotlin method \texttt{getX()} so a \textbf{Call} relation is extracted with Java source and Kotlin target information.

\vspace{-3mm}
\section{Evaluation}

\textbf{Subjects:}  Our subjects are {\sffamily DSU-Sideloader} (1259 Stars, 94.1\% Kotlin, 4.5\% Java, 7k LOC), {\sffamily Flap} (286 Stars, 52.4\% Kotlin, 47.6\% Java, 16k LOC) and {\sffamily RootEncoder} (738 Stars, 55.8\% Kotlin, 42.1\% Java, 51k LOC). We chose these projects because they have a high number of stars and range in size from 7k to 51k LOC, allowing us to test our tool's performance. Additionally, the Kotlin ratio varies among these three projects, from 94.1\% in {\sffamily DSU-Sideloader} to 55.8\% in {\sffamily RootEncoder}, which helps us test \textit{Depends-Kotlin}'s ability to handle Kotlin-Kotlin and Kotlin-Java dependencies. 

\textbf{Run \textit{Depends-Kotlin}:} Following  Figure~\ref{fig:framework}'s architecture, \textit{Depends-Kotlin} was implemented in Kotlin (the new Kotlin-Parser module) and Java (the legacy Depends-Core and Java-Parser modules). The README in the provided GitHub link specifies how to build and run this tool. Essentially, it takes the source file folder as input and outputs a JSON file with entities and dependency relations.

\textbf{Results:} Table~\ref{tbl:extracted-deps} presents the extracted dependencies by \textit{Depends-Kotlin}. The 1st column shows the dependency relations supported by our tool. The 2nd columns list the number of extracted dependencies in {\sffamily DSU-Sideloader}, with each sub-column representing dependency relations from a particular source (Java or Kotlin) to a particular destination (Java or Kotlin). For example, 19 in the K-J column under {\sffamily DSU-Sideloader} and the {\sffamily Import} row means there are 19 instances where Kotlin files import a Java class, enum, or static method. 423 in the J-K column under {\sffamily RootEncoder} and the {\sffamily Call} row means there are 423 instances where an expression in Java code invokes a Kotlin method in {\sffamily RootEncoder}. 

Table~\ref{tbl:extracted-deps} shows that all 13 dependency relations can be extracted from these three projects, with {\sffamily Use}, {\sffamily Call}, and {\sffamily Import} being the top three most frequent. We also observed that {\sffamily Delegate} and {\sffamily Extension} have only \colorbox{gray!30}{2} and \colorbox{gray!30}{9} instances, respectively, indicating that these two new syntax features are not widely used in these three projects. The last row shows the total number of four kinds of relations in each project. There are \colorbox{gray!30}{1476} K-K and \colorbox{gray!30}{309} J-J relations in {\sffamily DSU-Sideloader} with 94.1\% Kotlin code, compared to \colorbox{gray!30}{6990} K-K and \colorbox{gray!30}{6218} J-J relations in {\sffamily RootEncoder} with 55.8\% Kotlin code. It makes sense that a higher ratio of Kotlin code tends to result in a higher ratio of K-K relations in a project. We also observed a high ratio of K-J (\colorbox{orange!50}{420}) and J-K (\colorbox{green!40}{1307}) interactions in {\sffamily RootEncoder} and significant K-J relation instances in {\sffamily DSU-Sideloader} (\colorbox{orange!50}{152}) and {\sffamily Flap} (\colorbox{orange!50}{27}), which demonstrates our tool's ability to capture cross-language dependencies. Figure~\ref{fig:visual} presents the J-K (green arrow) and K-J (orange arrow) dependencies in the three subjects, with each node denoting a source file. As we can see, K-J and J-K dependencies are not limited to specific interfaces, demonstrating good interoperability between Java and Kotlin.

\vspace{-5mm}

\subsection{Accuracy Verification}

To our knowledge, there is no available tool to directly extract Kotlin(-Java) dependency relations from source code. Although CodeQL (\url{https://codeql.github.com}) and DependencyPlugin (\url{https://github.com/autonomousapps/dependency-analysis-gradle-plugin}) can extract dependencies from Kotlin projects, they require compiling the entire project or only provide package-level dependencies, making them unsuitable for comparison with our tool.

\textbf{Compiler Reference Check:} We leveraged JetBrains' Program Structure Interface (PSI) to conduct a first-round accuracy check. PSI is part of the Java/Kotlin compiler and can find references between elements in programs. If an extracted dependency instance by our tool can be found in references between two elements in PSI, we labeled it as ``\textit{Found}''; otherwise, we labeled it as ``\textit{NotFound}''. We calculate $Precision= \frac{Found\ Instances}{Found\ Instances\ + \ NotFound\ Instances}$ and $Recall= \frac{Found\ Instances}{Dependency\ Instances\ in\ PSI\ References}$ . This is a coarse comparison, as PSI does not provide detailed dependency relations and we can only check the existence of extracted dependencies in PSI. Table~\ref{tbl:accuracy} presents the precision value in our three subjects. Our tool shows promise in resolving dependencies both within the same language and across languages. Specifically, the average precision of J-J and K-K dependencies are 98.2\% and 97.9\%, respectively, while the average precision of K-J and J-K dependencies are 90.0\% and 99.5\%, respectively. Generally, the average precision of 97.9\% indicates that the capture dependencies by \textit{Depends-Kotlin} is precise. Due to space limit, the recall table is provided in the dataset link (see the abstract). The average recall across the three subjects is 85.0\%, indicating that 85.0\% of the dependencies in the ground truth can be correctly captured by \textit{Depends-Kotlin}.

\textbf{Manual Check:} For each dependency type, we also randomly selected 5 instances from J-J, K-K, J-K, and K-J (if available) in our three subjects, and two authors with four years of Java/Kotlin experience conducted a thorough examination of the instances (the dataset is shared in the GitHub link provided in the abstract). They built the projects in IntelliJ IDEA and independently traced entities to check dependency instances. After completing this, they discussed the results until an agreement was reached. The manual inspection results served as ground truth and were compared with the dependency instances generated by \textit{Depends-Kotlin}. The comparison shows that our tool can correctly capture 206 out of 213 dependencies (96.7\% precision).

In general, both the compiler reference check and the manual check show that \textit{Depends-Kotlin} can achieve good accuracy in capturing dependency relations.

\begin{table}[!tbp]
    \newcommand{\onehundred}{{\cellcolor{darkgray!30}100.0\%}}
    \centering
    \small
    \caption{Precision Value of Three Projects}
    \label{tbl:accuracy}
    \vspace{-4mm}
    \begin{tabular}{c|c|c|c|c|c}
    \hline
        & K-J    & J-K & K-K & J-J & \textbf{Average} \\ \hline
        Import   & \onehundred{} & \onehundred{} & \onehundred{} & \onehundred{} & \onehundred{} \\ \hline
        Contain  & \onehundred{} & \cellcolor{gray!30}99.6\% & \cellcolor{gray!30}99.4\% & \onehundred{} & \cellcolor{gray!30}99.6\% \\ \hline
        Extend   & \onehundred{} & \onehundred{} & \onehundred{} & \onehundred{} & \onehundred{} \\ \hline
        Implement & \onehundred{} & \onehundred{} & \onehundred{} & \onehundred{} & \onehundred{} \\ \hline
        Call & \cellcolor{lightgray!30}89.5\% & \cellcolor{gray!30}98.8\% &\cellcolor{gray!30} 98.1\% & \cellcolor{gray!30}96.7\% & \cellcolor{gray!30}97.4\% \\ \hline
        Create   & \onehundred{} & \onehundred{} & \cellcolor{gray!30}98.2\% & \cellcolor{gray!30}95.2\% & \cellcolor{gray!30}98.2\% \\ \hline
        Cast     & \onehundred{} & \onehundred{} & \onehundred{} & \onehundred{} & \onehundred{} \\ \hline
        Annotation & - & - & - & \onehundred{} & \onehundred{} \\ \hline
        Use      & \cellcolor{lightgray!30}78.2\% & \cellcolor{gray!30}99.5\% & \cellcolor{gray!30}96.7\% & \cellcolor{gray!30}98.7\% &\cellcolor{gray!30} 97.4\% \\ \hline
        Parameter & \onehundred{} & \onehundred{} & \cellcolor{gray!30}99.7\% & \cellcolor{gray!30} 99.5\% & \cellcolor{gray!30}99.7\% \\ \hline
        Return   & \onehundred{} & \onehundred{} & \onehundred{} & \cellcolor{lightgray!30}85.7\% & \cellcolor{gray!30}97.9\% \\ \hline
        Delegate & - & - & \onehundred{} & - & \onehundred{} \\ \hline
        Extension & - & - & \cellcolor{lightgray!30}88.9\% & - & \cellcolor{lightgray!30}88.9\% \\ \hline \hline
        \textbf{Average} & \cellcolor{gray!30}90.0\% & \cellcolor{gray!30}99.5\% & \cellcolor{gray!30}97.9\% & \cellcolor{gray!30}98.2\% & \cellcolor{gray!30}97.9\% \\ \hline
    \end{tabular}
    \vspace{-5mm}
\end{table}

\vspace{-2mm}
\subsection{Performance Evaluation}

Our experiments were conducted on a computer (AMD Ryzen 7 4800H @ 2.9GHz, 8GB RAM), and the performance of our analysis is shown in Table~\ref{tbl:performance}. The whole dependency extraction consists of four stages: \textit{Source File Parsing}, \textit{Entity Extraction}, \textit{Dependency Relation Extraction}, and \textit{Result Output}. We applied instrumentation in the program and calculated each stage's running time. Due to our \textit{Entity Extraction} process and \textit{Antlr}'s \textit{Source File Parsing} occurring simultaneously, it is difficult to split \textit{Source File Parsing} and \textit{Entity Extraction}. We leveraged the IntelliJ Profiler to assess the time ratio consumed by \textit{Source File Parsing}, then multiplied this ratio by the total running time of the two stages. As shown in Table~\ref{tbl:performance},  for project sizes of 7k LOC and 51k LOC, the analysis time ranges from 32.0 seconds to 114.9 seconds. The main functions of this tool —\textit{Entity Extraction} and \textit{Relation Resolution}—consumes a reasonable amount of time. The most time consuming stage is \textit{Source File Parsing}, which is handled by \textit{Antlr}, taking 87.8\%, 77.8\% and 78.8\% in {\sffamily DSU-Sideloader}, {\sffamily Flap} and {\sffamily RootEncoder}'s total analysis time. 
\begin{table}[!tbp]
    \centering
    \small
    \caption{Performance of the Four Stages in Three Subjects}
    \label{tbl:performance}
    \vspace{-4mm}
    \begin{tabular}{|c|c|c|c|}
    \hline
         & \textbf{DSU-Sideloader} & \textbf{Flap} & \textbf{RootEncoder}  \\ \hline
        \textit{Source File Parsing} & 28.1s  & 28.8s & 90.5s \\ \hline
        \textit{Entity Extraction} & 0.3s  & 1.1s & 3.9s \\ \hline
        \textit{Relation Resolution} & 3.3s  & 6.5s & 19.8s \\ \hline
        \textit{Result Output} & 0.3s  & 0.6s & 0.7s \\ \hline \hline
        \textbf{Total} & 32.0s  & 37.0s  & 114.9s  \\ \hline
    \end{tabular}
    \vspace{-4mm}
\end{table}

%Our future work will focus on improving this part, with directions including modifying the Kotlin grammar rules and adopting other parsers with better performance.
\vspace{-3mm}
\section{Conclusions and Future Work}

This paper proposes the \textit{Depends-Kotlin} tool, which can extract dependency relations in Kotlin-Java projects. The evaluation results from three subjects show that \textit{Depends-Kotlin} can capture Kotlin-Kotlin and Kotlin-Java dependencies accurately and efficiently. With cross-language development gaining popularity~\cite{li2021multi,mathew:2021cross,el:2024onespace}, code changes in one language can easily propagate to and impact other languages. Our tool has the potential to assist developers in handling cross-language scenarios, such as Kotlin-Java code smell detection, code migration, and architecture analysis of such systems.

The current version of \textit{Depends-Kotlin} has several limitations, such as the inability to resolve function calls on variables of generic types or determine the type of variables in lambda expressions (which results in a 85.0\% recall), as well as a slow performance for source code parsing by \textit{Anltr}. We plan to address these limitations in future work, with directions including modifying the Kotlin grammar rules and adopting other parsers with better performance.

%Our future work will focus on improving these part, with directions including modifying the Kotlin grammar rules and adopting other parsers with better performance.

%Currently, Depends-Kotlin cannot resolve function calls on variables of generic types or determine the type of variables in lambda expressions. We plan to address these limitations in future work.
%Furthermore, the plug-in architecture allows  to be extendable to other cross-languages. 

%Our future work plans to focus on improving its performance on large-scale projects.

% the analysis of links among different languages is essential for such software systems. To our knowledge, our prototype is the first to detect Kotlin-Java dependency relations based on static code syntax analysis.

%show that \textit{Depends-Kotlin} can achieve good performance on small or medium-sized projects but not very well on large projects. 

%This can also aid in understanding the software systems with the interactions between different programming languages.

%A comparison with a confirmed benchmark demonstrates \textit{Depends-Kotlin}'s accuracy in capturing dependency relations and invocations in Kotlin(-Java) code.

% \newpage
% \balance
\bibliographystyle{ACM-Reference-Format}
\bibliography{reference}

\end{document}